\documentclass[12pt]{article}
\usepackage{epsf,epsfig}
\def\beq{\begin{equation}}
\def\eeq{\end{equation}}
\def\beqar{\begin{eqnarray}}
\def\eeqar{\end{eqnarray}}
\def\ga{\mathrel{\raise.3ex\hbox{$>$\kern-.75em\lower1ex\hbox{$\sim$}}}}
\def\la{\mathrel{\raise.3ex\hbox{$<$\kern-.75em\lower1ex\hbox{$\sim$}}}}
\def\mg{m_{3/2}}

\begin{document}

\begin{titlepage}
\pagestyle{empty}
\baselineskip=21pt
\rightline{hep-ph/0107212}
\rightline{UMN--TH--2017/01, TPI--MINN--01/33}
\vskip 0.35in
\begin{center}
{\large{\bf The Inflatino Problem in Supergravity Inflationary Models
}}
\end{center}
\begin{center}
\vskip 0.05in
{{\bf Hans Peter Nilles}$^1$, 
{\bf Keith A.~Olive}$^{2}$
 and {\bf Marco Peloso}$^1$
\vskip 0.05in
{\it
$^1${Physikalisches Institut, Universitat Bonn \\
Bonn, Germany, Nusallee 12, D53115}\\
$^2${Theoretical Physics Institute, School of Physics and Astronomy,\\
University of Minnesota, Minneapolis, MN 55455, USA}\\
}}
\vskip 0.35in
{\bf Abstract}
\end{center}
\baselineskip=18pt \noindent
We consider the potential problems due to the production of inflatinos and
gravitinos after inflation. Inflationary models with a single scale set by
the microwave background anisotropies have a low enough reheat temperature
to avoid problems with the thermal production of gravitinos. Moreover, the
nonthermal production of gravitinos has been shown to be sufficiently
small if the sector ultimately responsible for supersymmetry breaking is
coupled only gravitationally to the inflationary sector. Still, in some
models, inflatinos can be created during preheating with a substantial
abundance. The main contribution to the gravitino abundance may thus come
from their decay into the inflaton, or into its scalar partner, as well as
from the inverse processes. We show that this production needs to be
strongly suppressed. This suppression can be realized in the simplest
scenarios which typically have a sufficiently high inflationary scale.
\vfill
\vskip 0.15in
\leftline{July 2001}
\end{titlepage}
\baselineskip=18pt

\section{Introduction}

It is well known that in the particle spectrum of the minimal
supersymmetric  standard model (the MSSM with R-parity conservation), the
lightest stable particle (LSP) is an excellent cosmological dark matter
candidate~\cite{hg,EHNOS}. Indeed, in the constrained version of the MSSM
based on minimal supergravity, one can easily satisfy all of the recent
accelerator constraints, and at the same time have an LSP with an
acceptable and significant cosmological relic density \cite{relic}. 
However, it is also well known that in these same models, there is  a
potential problem with the abundance of gravitinos in the early Universe
\cite{grprob}. 

Inflationary models are also subject to potential gravitino problems. If
the Universe reheats to sufficiently high temperatures, gravitinos will be
produced thermally with excessive abundances \cite{EHNOS,nos,kl}. For a
gravitational decay of the inflaton field the reheat temperature is
directly related to the inflationary scale. When this scale is fixed to
match the observed size of density fluctuations, the reheat temperature is
typically low enough to avoid thermal overproduction. Recently, the
non-thermal production of gravitinos has also been considered~\cite{gra1}.
It has been shown~\cite{gra2} that in this case too, gravitinos are
produced with a sufficiently low abundance, provided the inflationary
sector and the one responsible for supersymmetry breaking are distinct
and only gravitationally coupled.

The nonthermal production is much more relevant for the fermionic partner
of the inflaton, the inflatino, which in some cases can be produced with a
significant abundance. Thus, gravitinos may be overproduced through
inflatino decay, if the channel inflatino $\rightarrow$ inflaton (or its
scalar partner)
$+$ gravitino is kinematically allowed. Depending on the relative masses
of the inflaton and inflatino, significant production may be expected
instead by the inverse process. This is particularly true if the inflaton
sector is only gravitationally coupled to matter, since in this case the
above decays will have a rate comparable to the one generating the
thermal bath. Indeed, in such a scenario,  the decay channels into
gravitinos need to be strongly suppressed. We discuss the possible
kinematic suppression of this channel and derive a strong upper bound on
the scale of inflation. We further show that this bound is satisfied by
the simple single scale (supergravity) inflationary models.

We begin by reviewing the gravitino problem with respect to both thermal
and non-thermal production. We focus on simple models of inflation with a
single scale set by the microwave background anisotropies. In section 3,
we formulate the inflatino problem and discuss solutions for both the
gravitational and non-gravitational decay of the inflaton. This analysis
is generalized in section 4, where we discuss the merits of lowering the
scale of inflation. In section 5, we consider two specific models of
inflation: 1) a model of new inflation based on supergravity, and 2)
chaotic inflationary models. A discussion and concluding remarks are made
in section 6.

\section{The Gravitino Problem} \label{sec2}

Unless very massive, $\mg \ga 20$ TeV, gravitinos would disrupt the
successful predictions of big bang nucleosynthesis. The argument is
relatively simple: because of their late gravitational decays, gravitinos
dominate the energy density of the Universe which becomes matter dominated
with a Hubble expansion rate given by $H \sim \mg^{1/2} T^{3/2} /M_P $,
where $M_P$ is the (reduced) Planck mass. Gravitino decay occurs when the
decay rate $\Gamma_{3/2} \sim \mg^3/M_P^2 \sim H$ or at $T_D \sim
\mg^{5/3}/M_P^{2/3}$. Subsequently, the Universe is reheated to a
temperature  $T_R \sim  \rho(T_D)^{1/4} \sim m_{3/2}^{3/2}/M_P^{1/2}$. The
limit on the gravitino mass is obtained by requiring $T_R \ga 1$ MeV, thus
allowing big bang nucleosynthesis to occur in a radiation dominated
Universe.

However, since the gravitino abundance is diluted during
inflation \cite{grinf}, this simple constraint is significantly altered.
In fact, the constraint should be expressed as a function of the gravitino
mass and abundance after {\em inflationary} reheating.  
The most restrictive bound on their number density comes form the
photo-destruction of the light elements produced during nucleosynthesis
\cite{grnuc}
\beq
	n_{3/2}/s  \la 10^{-13}   (100 {\rm GeV}/m_{3/2}   )		
\label{bound}
\eeq
for lifetimes $ > 10^4$  sec.  The thermal production of gravitinos regenerated after
inflation has been estimated \cite{EHNOS,nos,grinf,grnuc}
\beq
n_{3/2}/s \sim (\Gamma /H)(T_{3/2}/T_\gamma)^3 \sim \alpha N(T_R) (T_R
/M_P )(T_{3/2}/T_\gamma)^3	
\label{bound1}
\eeq
where $\Gamma \sim \alpha N(T_R) (T_R^3/M_P^2)$  is the production rate of
gravitinos, N is the number of degrees of freedom, and the ratio
$(T_{3/2}/T_\gamma)^3$ accounts for the dilution of gravitinos by the
annihilations of particles between $T_R$ and nucleosynthesis. From the
gravitino regeneration rate one can derive a bound on $T_R$
\beq
	T_R \la  10^9~{\rm GeV} (100~{\rm GeV}/m_{3/2})		
\label{grlimit}
\eeq
For recent discussions on this bound, see~\cite{boundtr}.

The significance of the above constraint depends on the model of
inflation. We assume for simplicity that inflation is governed by a single
scale such that the inflaton potential is of the form $ V( \phi ) =
\Delta^4 P( \phi/M_P )$, where $P(\phi/M_P)$ is some suitable function
which generates inflation. The scale $\Delta$ can be related to the size
of large scale fluctuations measured by COBE~\cite{cobe}. In general this
relation is dependent in the choice of $P\,$, so that different values for
$\Delta$ can be considered. However, several of the proposed models
typically give~\cite{infrev}
\beq
{\delta \rho \over \rho} \simeq {H^2 \over 10 \pi^{3/2} \dot{\phi}}
\simeq O(100) {\Delta^2 \over {M_P}^2}
\label{perts}
\eeq
which, once related to the COBE results, implies~\cite{infl}
\begin{equation}
{\frac{\Delta^2}{M_P^2} = {\rm few} \times{10^{-8}}}
\label{cobedel}
\end{equation}

Fixing $({\Delta^2}/{M_P^2})$ has several general consequences for
inflation~\cite{eeno}. For example, the Hubble parameter during inflation
becomes ${{H^2} \simeq ({\Delta^4}/{3 M_P^2})}$, leading to $H \sim  
10^{-7}M_P$. The duration of inflation is $\tau \simeq
{M_P^3}/{\Delta^4}$, and the number of e-foldings of expansion is $H\tau
\sim({M_P^2}/{\Delta^2}) \sim 10^{9}$. If the inflaton decay rate goes as
$\Gamma \sim {m_{\phi}^3}/{M_P^2} \sim {\Delta^6}/{M_P^5}$, the Universe
recovers at a temperature $T_R \sim (\Gamma{M_P})^{1/2} \sim
{\Delta^3}/{M_P^2} \sim 10^{-11} {M_P} \sim 10^8$ GeV. This low reheating
temperature appears to be safe with regards to the gravitino limit
(\ref{grlimit}) discussed above.  

Recently, considerable attention has been focused on the possible
non-thermal production of gravitinos after inflation~\cite{gra1,gra2}.
Fermionic quanta can be created~\cite{gprt} in significant amounts during
the first stages of reheating (preheating). When applied to supergravity
models, preheating will lead to the non-perturbative production of the
fermionic partner of the inflaton, the inflatino, and in general, of any
other fermion which is strongly coupled to the inflaton field. If, for
example, there is substantial mixing between the inflatino and the
longitudinal component of the gravitino, the goldstino, preheating may
result in an overproduction of gravitinos. As is well known, the goldstino
is a linear combination of the fermionic partners of the scalars
responsible for supersymmetry breaking. During inflation and the beginning
of reheating, supersymmetry is mainly broken by the inflaton implying a
strong correspondence between the inflatino and goldstino at this early
stage. However, this correspondence does not necessarily hold at late
times, since supersymmetry may be  broken by other fields in the true
vacuum of the theory. In this case, the final gravitino abundance will be
much smaller than the inflatino abundance \cite{gra2}. Indeed, this is the
case in most models as it is natural to distinguish between inflation and
supersymmetry breaking due to the very different energy scales associated
with the two phenomena. The relic abundance of gravitinos will thus
ultimately be related to the strength of the coupling between these two
sectors in a given model.

If the scale of inflation, $\Delta$, is much larger than the other scales
in the theory, the nonthermal production of inflatinos can be accurately
computed by just considering the inflationary sector. Typically, particle
creation occurs during the very first few oscillations of the inflaton on
a timescale which is the inverse of the inflaton mass, $m_\phi \sim
\Delta^2/M_P\,$. Most of the inflatinos are produced with momenta $k \la
m_\phi\,$, and hence their initial number density is approximately given
by~\cite{gra1}
\begin{equation}
n_{\tilde \phi} \sim 0.01 \, m_\phi^3 \sim 0.01 \left(
\frac{\Delta^2}{M_P}
\right)^3
\end{equation}
The final inflatino abundance is $Y_{\tilde \phi} \equiv n_{\tilde \phi}/ s $
where $s \sim \rho^{3/4}$ is the entropy density at the time of
inflaton decay. For now, let us generalize our previous consideration of 
inflaton decay, and assume only that the decay is given by $\Gamma$
(rather than assuming specifically a gravitational decay).  In this case,
if inflaton oscillations begin when the scale factor is $R=R_\phi$, the
energy density in oscillations is $\rho_\phi \simeq m_\phi^2 M_P^2
(R_\phi/R)^3$ and $H \simeq m_\phi (R_\phi/R)^{3/2}$. Decays occur when
$H \simeq
\Gamma$, or when $R = R_{d \phi} = (m_\phi/\Gamma)^{2/3} R_\phi$.
During inflaton oscillations, the inflatino number density also scales as
$R^{-3}$ so that at the time of decay, $n_{\tilde \phi} \sim 0.01 \,
m_\phi^3 (R_\phi/R_{d \phi})^3 \sim 0.01 m_\phi \Gamma^2 \sim 0.01
{\Delta^2} \Gamma^2 / {M_P} $.  The entropy density is also easily
computed at decay, $ s \sim (\Gamma^2 M_P^2)^{3/4} \sim T_R^3$. 
Thus for a massive inflaton, one finds
\begin{equation}
Y_{\tilde \phi} \sim 0.01 \, \left( \frac{\Delta^2
\Gamma^{1/2}}{M_P^{5/2}}
\right) \sim 0.01 \, \left( \frac{\Delta}{M_P}
\right)^3
\,
\left(
\frac{T_R}{\Delta} \right)
\label{inflab}
\end{equation}
We see, therefore, that a late inflaton decay (low $T_R$) results
in a small inflatino abundance. This is due to the fact that the energy
density stored in the coherent oscillations of a massive inflaton
redshifts as matter ($\rho \sim R^{-3}$), so that the quantity $n_{\tilde
\phi}/\rho^{3/4}$ decreases with time. 

For the case of a gravitational decay, if we take $\Delta$ from
eq.~(\ref{cobedel}) and its corresponding reheat temperature $T_R \sim
10^8$ GeV, we find a negligible inflatino abundance $Y_{\tilde \phi} \la
10^{-19}$. For a massless inflaton the final abundance will be
significantly larger, due to the stronger decrease of the inflaton energy
density. For example, for a $V \sim \phi^4$ potential, the energy density
of $\phi$ redshifts as radiation, and the last factor $T_{R}/\Delta$ is
absent in the final abundance~(\ref{inflab}). In this case, the final
inflatino abundance may be as large as  $Y_{\tilde \phi} \simeq 10^{-12}$.

The calculation of the nonthermal production of the longitudinal gravitino
component in models with more than just the inflationary
sector~\cite{gra2} is more complicated. The mixed inflatino--gravitino
system is very involved, and analytical solutions for the gravitino
abundance are still lacking. A consistent quantization of the system, with
an accurate definition of the occupation numbers of the fermionic
eigenstates, is however available. For the case in which the sectors
responsible for inflation and supersymmetry breaking are coupled only
gravitationally, numerical computations show that  gravitino production is
restricted to a safe level. The longitudinal gravitinos are created on a
physical timescale of the order $m_{3/2}^{-1}\,$, with a physical typical
momentum comparable with $m_{3/2} \ll \Delta$.

\section{The Inflatino problem}

In the previous section, we have shown that if we restrict the
inflationary model so that it is coupled only gravitationally to the other
sectors of the theory, both the thermal and the nonthermal production of
gravitinos are reduced to a safe level. We have also seen that the
nonthermal production of inflatinos is model dependent and can be
substantial. If this is the case, and if the inflatino is heavier than the
inflaton, the subsequent gravitational decays of the inflatino (to an
inflaton and gravitino) could lead to an overproduction of gravitinos
(other decays of the inflatino in some specific models are discussed
in~\cite{abm}).

On the contrary, when the inflaton is heavier than the inflatino, then
the decay  channel $\phi \to {\tilde \phi} + {\tilde G}$ may be
problematic if kinematically allowed \cite{nos}.  The density of
gravitinos produced by inflaton decays is easily estimated.  By comparing
the number density of inflatons just prior to decay, $n_\phi \sim m_\phi
M_P^2 (R_\phi/R_{d \phi})^3 \sim m_\phi^5/M_P^2$ to the entropy density
just after decay $s \sim m_\phi^{9/2} / M_P^{3/2}$, one finds $n_\phi/s
\sim (m_\phi/M_P)^{1/2} \sim \Delta/M_P$. (It is also convenient to write
$n_\phi/s \sim (\Gamma M_P)^{1/2}/m_\phi \sim T_R/m_\phi$).  If the
branching fraction for inflaton decays to ${\tilde \phi} + {\tilde G}$ is
$1/N$, then the resulting gravitino abundance is
\beq
Y_{3/2} = \frac{n_{3/2}}{s} \sim \frac{\Delta}{N M_P} \sim 10^{-6}
\label{decay1}
\eeq
in clear violation of the bound (\ref{bound}). This is what we call the
inflatino problem. We note that, for a purely gravitational inflaton
decay, the limit from direct production is stronger than that due to
thermal production. Indeed, by requiring $Y_{3/2} \la 10^{-13}$,
eq.~(\ref{decay1}) yields the constraint $\Delta \la 10^{-11} M_P$ on the
inflationary scale, which is stronger than the bound from the thermal
production of gravitinos. Requiring $T_R \la 10^9$ GeV, and noting that
for gravitational decays, $T_R \sim (\Gamma M_P)^{1/2} \sim \Delta^3
/M_P^2$, one finds $\Delta \la 10^{15}$ GeV. 

One may hope to relax this difficulty by allowing for non-gravitational
couplings of the inflaton to matter.  For example, if we suppose that
the decay rate to matter is given by $\Gamma_m \sim g^2 m_\phi$, then the
gravitino abundance will be suppressed, $Y_{3/2} \sim (\Gamma_G /
\Gamma_m) n_\phi/s$ where $\Gamma_G \sim m_\phi^3/M_P^2$ is the
gravitational decay rate.  In this case, decays occur earlier and
$n_\phi/s \sim g (M_P / m_\phi)^{1/2} \sim g M_P / \Delta$.  The
gravitino abundance produced by inflaton decays is now
\beq
Y_{3/2} \sim  \frac{\Delta^3}{g N M_P^3}
\label{decay2}
\eeq
Therefore, the constraint on $\Delta$ from the gravitino abundance becomes
$\Delta \la \, {\rm few} \times \, 10^{-4} g^{1/3}M_P$. However, for
stronger than gravitational decays ($g > \Delta^2/M_P^2$), the constraint
from the thermal production of gravitinos becomes more significant. From
$T_R \sim g \Delta\,$, one finds the limit $\Delta \la 10^9$ GeV$/g\,$.
The combination of the two bounds from the thermal and the direct
gravitino production is weakest when $g \sim 10^{-\,15/4}\,$. For this
value, the limit $\Delta \la \, \times 10^{-5} M_P$ is found.  Thus the
simple generic scenario (with a single scale set by the CMB) is not
possible if the decay of the inflaton to inflatino $+$ gravitino occurs
unimpeded.

Of course, as noted in \cite{nos}, if it should happen that the decay is
kinematically forbidden $(|m_\phi - m_{\tilde \phi}| < \mg$), then the
simple single scale inflationary model works very well. Here, we note that
if the scale of supersymmetry breaking, $\mu$, is significantly below that
of inflation, i.e., $\mu \ll \Delta$, then even though the decay $\phi \to
{\tilde \phi} + {\tilde G}$ is allowed, it will be naturally kinematically
suppressed with respect to the other decay channels of the inflaton field.
This will open an allowed window for $\Delta$ even in the simplest
scenarios.

Independent of the details of the decay, the rate will always carry a
final state momentum suppression factor.  The overall decay rate can be
written as $\Gamma \sim (1/m_\phi) |{\cal M}|^2 (p/m_\phi)$, where
$|{\cal M}|$ is the amplitude for decay and the final state momentum
suppression factor is
\beq
2p/m_\phi = \left( 1 - \frac{2(m_{\tilde \phi}^2 + m_{3/2}^2)}{m_\phi^2} +
\frac{(m_{\tilde \phi}^2 - m_{3/2}^2)^2}{m_\phi^4}   \right)^{1/2}
\sim \frac{\mu^2}{\Delta^2}
\label{poverm}
\eeq
recalling that $m_\phi \sim m_{\tilde \phi} \sim \Delta^2/M_P$ and
$m_{3/2} \sim \mu^2/M_P$. Thus in models in which $\mu \ll \Delta$, there
will be a significant suppression in the production of gravitinos by
either inflaton or inflatino decay (note that additional suppression
may come from the specific form of the amplitude ${\cal M}$\, as well).

If we take into account the suppression factor~(\ref{poverm}) in the bound
for the direct gravitino production by inflaton decay, and we combine it
with the limit coming from the thermal production, we find that the scale
$\Delta$ must lay within the interval
\beq
10^{13} \frac{\mu^2}{N M_P} \la \Delta \la 10^{15} {\rm GeV}
\label{window}
\eeq
For $\mu \sim 10^{-8} M_P$, the {\em lower} bound in (\ref{window}) is about
$10^{13}$ GeV. For smaller values of $\Delta$ the kinematical suppression
factor~(\ref{poverm}) is no longer capable of maintaining a low gravitino
abundance. We notice that the scale $\Delta \sim 10^{-\,4} \, M_P$ as given in
eq.~(\ref{cobedel}) is within the allowed interval. Indeed, including the
suppression factor~(\ref{poverm}) to eq.~(\ref{decay1}), the gravitino
abundance produced by decays is now
\beq
Y_{3/2}  \sim \frac{\Delta}{N M_P} \frac{\mu^2}{\Delta^2} \sim
\frac{m_{3/2}}{N \Delta} \sim 10^{-14}
\label{decay3}
\eeq
and clearly satisfies the bound (\ref{bound}). The simplest models of
single scale inflation with purely gravitational decays and a scale given
by~(\ref{cobedel}), therefore, do not suffer from a gravitino problem.

For completeness, we  note that when non-gravitational decays are also
allowed, the kinematic suppression allows a wide range of values for
$\Delta$ (and $g$). While the constraint from the thermal production of
gravitinos is unchanged ($g \Delta < 10^9$ GeV), the constraint from
decays becomes $\Delta \mu^2 / g N M_P^3 < 10^{-13}$ which for $\mu =
10^{-8} M_P$ becomes $\Delta < 10^{23} g $ GeV. 

Before we move on to specific examples of inflationary models, we note
that there are at least two other potential sources of gravitino
production. First, there is the possibility for an additional contribution
to the direct production of gravitinos by the decay $\phi \to {\tilde G}
{\tilde G}\,$. During inflaton oscillations, the rate for such decays has
been estimated to be $\Gamma_{3/2} \sim \Delta^{10}/M^9$ \cite{sr}. This
implies that the abundance of gravitinos will be $Y_{3/2} \sim
(\Gamma_{3/2}/\Gamma) n_\phi/s \sim (\Delta/M_P)^5$. For all of the models
being discussed, this is sufficiently small. Second, the remaining scalar
degree of freedom in the inflaton supermultiplet may also be problematic
in certain cases. This field, the sinflaton $\phi'$, is normally described
by the complex direction of the full inflaton potential. The complex
direction is stable in most models, and thus  classically it is not
excited. However, if lighter than the inflatino, inflatino decay to the
sinflaton $+$ gravitino may yield another source of gravitino production.
The above arguments would therefore also apply to these decays as well. In
addition, it is possible that quantum fluctuations will excite the
sinflaton during inflation. Even though the sinflaton is far from being
massless, quantum fluctuations will lead to  $\phi'\sim H$ \cite{engo}.
However, in generic models since the amplitude of sinflaton oscillations
is much smaller than the corresponding amplitude for inflaton
oscillations, their decay into inflatinos is expected to be suppressed by
a factor of $(H/M_P)^2 \sim (\Delta/M_P)^4$.

\section{Lowering the scales of the inflationary sector}

As we have seen, the potential of the inflaton field during inflation is
constrained by the magnitude of density fluctuations measured by COBE. In
the slow roll regime ($3 \, H \, \dot{\phi} \simeq - \, V'$) eq.~(\ref{perts}) gives
\begin{equation}
V_H^{1/4} \simeq 0.027 \: \epsilon_H^{1/4} \: M_P
\label{cobeve}
\end{equation}
where $\epsilon^2 \equiv  M_p^2 \, \left( V'/V \right)^2 / 2 \ll 1$ is one
of the two slow-roll parameters (prime denoting derivative with respect to
$\phi$) and the suffix $H$ reminds us that the two quantities have to be
evaluated when the scales measured by COBE left the horizon, about $60$
e--foldings before the end of inflation.  As we have said, in the simplest
models of single scale inflation, this relation fixes the scale of the
inflaton potential $\Delta \equiv V_H^{1/4}$ to be about $10^{-\,4} \,
M_P\,$. However, models with a much smaller scale and acceptable density
fluctuations can be derived. As follows from eq.~(\ref{cobeve}), in models
with one single field this can be done at the expense of a small
$\epsilon$ parameter, that is by taking a very flat potential during
inflation. Such a flat potential may arise more naturally if more scalar
fields are present, as for example in hybrid inflationary
models~\cite{hybrid}. Due to this freedom, most of the above results of
the previous section have been given for an arbitrary inflationary scale.

In this section, $\Delta^4$ denotes the value of inflaton potential at the
end of inflation, when the reheating stage begins. Due to the slow motion
of $\phi$ during inflation, this scale is typically very close to $V_H\,$.
During reheating, the inflaton field oscillates about the minimum
$\phi_0$  of $V$, with $V \left( \phi_0 \right) = 0\,$. As is typical for
a massive inflaton, we assume that the quadratic term dominates the Taylor
expansion of $V$ around $\phi_0\,$.  We denote by $F$ the amplitude of the
inflaton oscillations at the initial time $t \sim H^{-\,1} \sim M_P /
\Delta^2\,$. While in the above discussions the natural assumption $F =
M_P$ (originally dubbed primordial inflation \cite{prim}) was made, in the
following we discuss how the results of the previous section are affected
when the value of $F$ is lower.

This analysis is particularly simplified by noticing that most of the
bounds previously discussed are directly related to the inflaton mass
$m_\phi$ rather than to the scale $\Delta\,$. For a quadratic potential
and generic values of $\Delta$ and $F$ one has $m_\phi \simeq \Delta^2 /
F\,$.

Let us first consider the case in which the inflaton decays only
gravitationally. For a quadratic potential, the reheating temperature is
given by $T_R \simeq \left( \Gamma \, M_P \right)^{1/2} \simeq
\sqrt{m_\phi^3/M_P}\,$, so that the thermal bound~(\ref{grlimit}) simply
gives
\begin{equation}
m_\phi \la 10^{12} \: {\rm GeV}
\label{fbound1}
\end{equation}

Following the same line of arguments of the previous sections, the
inflaton ``abundance'' at the decay time can be estimated to be
\begin{equation}
\frac{n_\phi}{\rho_\phi^{3/4}} \simeq \frac{T_R}{m_\phi}
\simeq \sqrt{\frac{m_\phi}{M_P}}
\label{finfla}
\end{equation}
The abundance of inflatinos produced nonthermally by the inflaton
oscillations is instead (also evaluated at the inflaton decay)
\begin{equation}
\frac{n_{\tilde \phi}}{\rho_\phi^{3/4}} \simeq 10^{-\,2} \, \frac{m_\phi
\, T_R}{M_P^2} \simeq 10^{-\,2} \left( \frac{m_\phi}{M_P} \right)^{5/2}
\end{equation}

We see that the inflaton abundance is always higher than the inflatino
abundance. Also the abundance of gravitinos produced nonthermally becomes
smaller as
$m_\phi$ decreases. Indeed, gravitinos will still be mainly produced at
the time $t \sim m_{3/2}^{-\,1}$ with a typical momentum $k \sim
m_{3/2}^{-\,1}\,$, independent of the value of $m_\phi\,$. However, a
lighter inflaton implies a longer lifetime $\tau_\phi\,$. Since the
quantity $n_{3/2}/\rho_\phi^{3/4}$ decreases in the time interval
$m_{3/2} < t < \tau_{\phi}\,$, lowering $m_\phi$ will thus decrease the
final nonthermal gravitino abundance.

Direct gravitino production still mainly occurs through the $\phi
\rightarrow {\tilde \phi} \, {\tilde G}\,$ decay, if kinematically
allowed. This production is reduced to a safe level as long as
\begin{equation}
m_\phi^{1/2} \ga \frac{10^{13}}{M_P^{1/2}} \, \frac{\mu^2}{N\,M_P}
\label{fbound2}
\end{equation}
where the kinematical suppression factor $m_{3/2}/m_\phi$ has been
included.

Rewriting eqs.~(\ref{fbound1}) and (\ref{fbound2}) in terms of $\Delta$
and $F$, the allowed window for $\Delta$ becomes
\beq
10^{13} \frac{\mu^2}{N M_P} \, \sqrt{\frac{F}{M_P}} \la \Delta \la
10^{15} {\rm GeV} \, \sqrt{\frac{F}{M_P}}
\label{fwin}
\eeq
For $F=M_P\,$, this coincides with the result~(\ref{window}) obtained in
the previous section.  For smaller $F\,$, smaller values of $\Delta$ must
be considered, as it is obvious from the scaling $m_\phi \simeq
\Delta^2/F\,$.

As before, additional freedom in the choice of parameters is allowed if the
inflaton has non-gravitational decays. For $\Gamma \sim g^2 \, m_\phi$ and
for $m_{3/2} \simeq 100 \,$ GeV, the bounds on the reheating temperature
and on the direct gravitino production become
\begin{equation}
m_\phi \la {\rm GeV} / g^2 \;\;,\;\;
m_\phi \la 10^{28} \, {\rm GeV} \, g^2
\end{equation}
Also in this case, the bounds can be written just in terms of the
inflaton mass. Notice that when the inflaton has nongravitational decays,
the limit from direct gravitino production gives a lower bound (rather
than an upper one) on $\Delta\,$. This follows from the different
scalings of the gravitational and the non gravitational rates with
$m_\phi\,$.

\section{Models}

Given the above arguments for the postinflationary production of gravitinos
through thermal and non-thermal effects, we will now apply them to some
specific examples.  First we consider a simple model of inflation based on N=1
supergravity \cite{nost,hrr}. We subsequently consider gravitino production in
models of chaotic inflation \cite{chaotic}. In both examples, we will assume a
superpotential of the form 
\begin{eqnarray}
W_T &=& W \left( \Phi \right) + \mu^2 \left( S + \beta \right) \nonumber\\
W \left( \Phi \right) &\equiv& \Delta^2 \, {\widetilde W} \left( \Phi \right)
\label{superp}
\end{eqnarray}
where we set $M_P=1$ for convenience.
The first term specifies the dynamics of the inflaton field $\phi$ (i.e. the
scalar component of $\Phi$) during inflation and the first stages of
reheating. The second term is known as the the Polonyi superpotential
\cite{polo}, and is the simplest example of gravitationally mediated
supersymmetry breaking involving only one chiral multiplet $S\,$.  We also
assume a minimal K\"ahler potential $G = K + \ln |W_T|^2$, with $K = \Phi^\dagger
\,
\Phi + S^\dagger
\, S$. In the minimum of the theory, the Polonyi scalar takes the vacuum
expectation value,
$\langle s
\rangle = \sqrt{3} - 1\,$, up to possible ${\hat \mu}^2 \equiv \mu^2/\Delta^2
\ll 1$ corrections which may arise from the interaction with the field
$\phi\,$. The parameter $\beta$ is fine-tuned to $2 - \sqrt{3}$ (again up to
possible ${\hat \mu}^2$ corrections) to cancel the cosmological constant.

In the following examples, we will assume that in the limit of $\mu \to 0$,
supersymmetry is unbroken, that is, supersymmetry is mainly (in some cases
completely) broken by the Polonyi field. Therefore, we have
$F_\phi \leq {\hat \mu}^2 F_s\,$, where, for a minimal K\"ahler potential, $F_i
\equiv e^{G/2} G_i = {\rm e}^{K/2} \, \left( \partial W_T / \partial \varphi_i
+ \varphi_i^*\, W_T \right) \,$ and $G_i = \partial G/\partial \varphi_i$. We
also have $W \left( \phi \right) \leq W \left( s
\right) {\hat \mu^2}\,$. From the specific form of the Polonyi superpotential
one finds $F_s = \sqrt{3} \mu^2 \,$, and the gravitino mass will be 
\begin{equation}
m_{3/2} \sim {\rm e}^{K/2} \,\mu^2
\label{mgra}
\end{equation}
These relations follow from the fact that in these models, 
${\widetilde W} \left( \phi \right)$ and
$F_\phi/\Delta^2$  are at most of order ${\hat
\mu}^4\,$ in the global  minimum of the theory.

With these assumptions, the real and the imaginary components of the Polonyi
field $s = \left( s_1 + i s_2 \right)/\sqrt{2}$ have masses
\begin{equation}
m_{s_1} \sim {\rm e}^{K/2} \, \mu^2 \sqrt{2 \, \sqrt{3}} \;\;\;,\;\;\;
m_{s_2} \sim {\rm e}^{K/2} \, \mu^2 \sqrt{4 - 2 \, \sqrt{3}}
\label{mpolo}
\end{equation}
Analogously, for the inflaton sector we decompose $\phi = \left( \phi_1 + i
\phi_2 \right) /\sqrt{2}$ and we denote the inflatino field by ${\tilde
\phi}\,$. Then, a simple calculation gives (assuming ${\widetilde W}' \leq
{\hat \mu}^2 \, {\widetilde W}'' \neq 0$ where prime denotes derivative with
respect to $\phi\,$)
\begin{eqnarray}
m_{\phi_1 \,,\, \phi_2} &=& m_{\tilde \phi} \pm m_{3/2} \,
\left( 1 - \frac{\sqrt{3}}{2} \right) + {\rm O} \, \left( {\hat
\mu}^4 \right) \nonumber\\
\frac{m_{\tilde \phi}}{\Delta^2} &=& {\rm e}^{K_0/2} \left[ {\widetilde W}'' -
{\hat \mu}^2 \, \left( \frac{{\widetilde W}'}{{\hat \mu^2}} \right)^2 + {\hat
\mu}^2 \, \frac{K_1}{2} \, {\widetilde W}'' \right] + {\rm O} \, \left( {\hat \mu}^4 \right)
\label{relat}
\end{eqnarray}
where $K_0$ and $K_1$ are the first two terms in the expansion of the function
$K = K_0 + {\hat \mu}^2 K_1 +\dots\,$.

Note that the masses of the inflatino and of the two inflaton components are
nearly degenerate, their difference  being related to the small gravitino mass
$m_{3/2}\,$. This is a consequence of having taken the supersymmetry
breaking scale
$\mu$ much smaller then the inflationary scale $\Delta\,$. We notice that the
specific choice of the Polonyi superpotential~(\ref{superp}) gives
$m_{\phi_2}
\la m_{\tilde \phi} \la m_{\phi_1}\,$, but that the mass differences are
always smaller than the gravitino mass $m_{3/2}\,$. Hence, the two
potentially dangerous  decay channels $\phi \rightarrow {\tilde \phi} +
{\tilde G}$ and ${\tilde \phi}
\rightarrow \phi + {\tilde G}$ are kinematically forbidden in these models.

It is important to stress that the above relations~(\ref{relat}) are
unaffected by the oscillations of the Polonyi field about its minimum. Indeed,
for $m_{3/2} \simeq 100\,$ GeV we have
\begin{equation}
\frac{\Gamma_{\rm grav}}{m_{3/2}} \sim 10^{-\,8} \, \left(
\frac{\Delta}{10^{-\,4}\,M_P} \right)^6
\label{splitok}
\end{equation}
where $\Gamma_{\rm grav} \sim m_\phi^3/M_P^2$ is the rate at which the
decay $\phi \rightarrow {\tilde \phi} + {\tilde G}$ would occur if not
kinematically suppressed or forbidden. The oscillations of the Polonyi
field start at the time $t=m_{3/2}^{-\,1}$ and are eventually damped by
the expansion of the Universe. For $t > m_{3/2}^{-\,1}\,$, their amplitude
scales inversely with time. Since the initial amplitude of the
oscillations and the value of the Polonyi field in the minimum are of the
same order ($M_P$), eq.~(\ref{splitok}) shows that these oscillations can
be soon neglected in the calculation of the mass spectrum.

Actually, in the following we will ignore the evolution of the Polonyi
field. This evolution, however, leads to a well known problem in cosmology
which is compounded in more modern theories collectively called the moduli
problem.~\cite{pp}

\subsection{New inflation}

A workable example of inflation is provided by~\cite{hrr}
\begin{equation}
W \left( \Phi \right) = \frac{\Delta^2}{2 \, M_P} \, \left( \Phi - \phi_0
\right)^2
\label{superp2}
\end{equation}
which is clearly of the form advocated in section 2.

During inflation and the first stages of reheating, the Polonyi scalar is
frozen at $s =0\,$. Moreover, since $\mu \ll \Delta\,$, the Polonyi sector is
completely negligible at this stage. Thus, for the moment, we can set
$\mu=0\,$ and decompose the inflaton field into its real plus
imaginary components,
$\phi = \left( \phi_1 + i \phi_2 \right)/\sqrt{2} \,$. Inflation occurs while
$\phi_1 \simeq 0\,$. Near the origin, the real direction is
relatively flat, the imaginary component has large positive mass, and it
is rapidly driven to zero. At the classical level, we can set $\phi_2
\equiv 0\,$, since $V'' $ is always positive in the imaginary direction.
Doing so, one finds
\begin{equation}
V = \frac{\Delta^4}{4} \, {\rm e}^{\phi_1^2/ \left( 2 \, M_P^2 \right)} \,
\left( \frac{\phi_1}{\sqrt{2} \, M_P} \, - 1 \right)^2 \, \left[ 1 + \sqrt{2}
\, \frac{\phi_1}{M_P} + \frac{\phi_1^2}{M_P^2} - \frac{\phi_1^3}{\sqrt{2} \,
M_P^3} + \frac{\phi_1^4}{4\,M_P^4} \right] \geq 0
\label{pot}
\end{equation}
where $\phi_0 = M_P$ has been set to have a vanishing cosmological constant in
the minimum at $\phi_1 = \sqrt{2} \, M_P\,$. Notice that in the vicinity
of the origin, both the linear and quadratic terms in $\phi_1$ cancel
when the exponential prefactor is expanded. Inflation in this model occurs
for
$\phi_1
\la M_P /
\left( 6
\,
\sqrt{2}
\right)
\,$. The inflationary scale
$\Delta$ which matches the microwave background anisotropies is about
$\Delta/M_P
\simeq 2.6 \times 10^{-\,4}\,$, and the spectral index is found to be $n_s
\sim 0.93\,$.

Let us now consider the late behavior of the fields and reintroduce the
dimensionless quantity ${\hat \mu} \equiv \mu / \Delta \ll 1\,$. The
presence of a nonvanishing Polonyi field slightly modifies the potential
for $\phi$ and the position of the minimum. Also, the nonvanishing vev of
$\phi$ forces a small modification to the Polonyi potential. These
effects, which are easily computed in an expansion series in ${\hat
\mu}^2\,$, must be taken into account in the computation of the spectrum
in the minimum of the theory. There is some freedom in the choice of the
parameters, since different combinations of $\phi_0$ and $\beta$ give a
zero cosmological constant. Up to order ${\hat
\mu^6}$ corrections, one finds
\begin{eqnarray}
\frac{\phi_0}{M_P} &=& 1 + f \, {\hat \mu^2} + g \, {\hat \mu}^4
\nonumber\\
\frac{\beta}{M_P} &=& \left( 2 - \sqrt{3} \right)  - {\hat \mu}^2 / 2 +
\left( \frac{3}{4} - \frac{\sqrt{3}}{2} - f \right) {\hat \mu}^4
\label{param}
\end{eqnarray}
with $f$ and $g$ arbitrary real numbers.

Decomposing $s = \left( s_1 + i s_2 \right) / \sqrt{2}\,$, one finds
$s_2 \equiv 0\,$, while the real components have their minima
at~\footnote{For these minima one has
\begin{eqnarray}
\frac{\partial V}{\partial \phi} = {\rm O } \, \left( {\hat \mu}^8 \right)
\;\;,\;\; \frac{\partial V}{\partial s} = {\rm O } \, \left( {\hat \mu}^{10}
\right) \;\;,\;\; V = {\rm O } \, \left( {\hat \mu}^{10} \right)
\nonumber
\end{eqnarray}}
\begin{eqnarray}
\frac{\langle \phi_1 \rangle}{\sqrt{2} \, M_P} &=& 1 + \left( f - 1 \right)
{\hat \mu}^2 + \left( g - f + \frac{1}{2} - \sqrt{3} \right) {\hat \mu}^4
\neq \phi_0 \nonumber\\
\frac{\langle s_1 \rangle}{\sqrt{2} \, M_P} &=& \sqrt{3} - 1 + \frac{{\hat
\mu}^2}{2} + \left( f + \frac{\sqrt{3}}{2} - \frac{9}{8} \right) {\hat \mu}^4
\end{eqnarray}
again up to order ${\hat \mu}^6$ corrections.

In the minimum, the $F-$terms are given by
\begin{equation}
\frac{F_s}{\Delta^2} \simeq \sqrt{3} \, {\hat \mu}^2 + \frac{\sqrt{3}}{2} \,
{\hat \mu}^4 \;\;\;,\;\;\; \frac{F_\phi}{\Delta^2} \simeq - \, \sqrt{3} \,
{\hat \mu^4}
\end{equation}
Hence, the inflaton sector provides a negligible (although non-vanishing)
contribution to supersymmetry breaking in the minimum of the theory. At
late times the goldstino can be identified with
the Polonyi fermion, up to small ${\hat \mu}^2$ corrections. The masses
of the gravitino and of the two components of the Polonyi field are given
by the general expressions~(\ref{mgra}) and (\ref{mpolo}), so that we have
\begin{equation}
{\hat \mu}^2 \simeq 2.8 \cdot 10^{-\,10} \, \left( \frac{m_{3/2}}{100 \:
{\rm  GeV}} \right)
\end{equation}

The general relations~(\ref{relat}) for the inflationary sector are also
respected. More explicitly, one finds
\begin{eqnarray}
m_{\tilde \phi} &=& \frac{\Delta^2}{M_P} \, {\rm e}^{\frac{5}{2} -\sqrt{3}}
\left[ 1 + \left( f + \frac{\sqrt{3}}{2} - \frac{5}{2} \right) {\hat \mu^2}
\right] \nonumber\\
m_{\phi_1} &=& \frac{\Delta^2}{M_P} \, {\rm e}^{\frac{5}{2} -\sqrt{3}}
\left[ 1 + \left( f - \frac{3}{2} \right) {\hat \mu}^2 \right]
\nonumber\\
m_{\phi_2} &=& \frac{\Delta^2}{M_P} \, {\rm e}^{\frac{5}{2} -\sqrt{3}}
\left[ 1 + \left( f - \frac{7}{2} + \sqrt{3} \right) {\hat \mu}^2 \right]
\end{eqnarray}
confirming that the two channels $\phi_1 \rightarrow {\tilde \phi} +
{\tilde G}$ and ${\tilde \phi} \rightarrow \phi_2 + {\tilde G}$ are indeed
kinematically forbidden.

With regard to the nonthermal production of inflatinos and gravitinos, we have
verified by numerical calculations that the model~(\ref{superp2}) gives
results in very good agreement with the ones obtained in~\cite{gra2} for the
chaotic inflationary superpotential $W \left( \Phi \right)= m_\phi^2 \, \Phi^2
/ 2 \,$ (see below). Inflatinos are produced during the first oscillations of
the inflaton field ($t \sim m_\phi^{-1} = M_P/\Delta^2$). Their final
abundance $Y_{\tilde \phi}$ is given in equation~(\ref{inflab}). As we have
remarked, the final value of $Y_{\tilde \phi}$ is typically very small, due to
the dilution of $n_{\tilde \phi}/\rho_\phi$ during the inflaton oscillations
in the massive inflaton case. The nonthermal production of gravitinos is
instead mainly due to the Polonyi oscillations, and indeed it occurs on a
timescale $t \sim m_s^{-1} \sim m_{3/2}^{-1} \,$. As we have already
mentioned, their final abundance is much smaller than that of inflatinos,
since they are typically produced with a physical momentum $k \sim m_{3/2}\,$.

\subsection{Chaotic inflation}

For completeness, we consider two types of chaotic models with
superpotentials
\begin{equation}
W \left( \Phi \right) = \frac{1}{2} \, m_\phi \, \Phi^2 \;\;\; {\rm and}
\;\;\; W \left( \Phi \right) = \frac{\lambda}{3} \, \Phi^3
\label{superp3}
\end{equation}
which for $\phi \ll M_P$ lead to the usual quadratic and quartic potentials
typical of chaotic inflation~\cite{chaotic}. As it is well known, the above
superpotentials do not lead to inflation in supergravity, since
corrections from the K\"ahler potential spoil the flatness of $V \left(
\phi \right)$ for
$\phi \ga M_P\,$ (at least in minimal supersgravity). One can still assume
that some corrections, relevant at high
$\phi\,$, will  generate a sufficiently flat potential so to render
inflation possible, and that the superpotentials~(\ref{superp3}) will be
accurate enough during the reheating stage (see  \cite{gl} for example,
for chaotic models fully consistent with supergravity). After inflation
the production of gravitinos can be very well estimated with the
superpotential~(\ref{superp3}), at $\phi \ll M_P\,$~\cite{gra2}.
However, the two parameters $m_\phi$ and $\lambda$ cannot be
directly linked to the size of the density fluctuations as in
eq.~(\ref{perts}), and the ``standard'' values $m_\phi \sim 10^{13}$ GeV
and  $\lambda \sim 10^{-\,13}$ should be taken only as indicative ones.
In both
cases, $\phi =0$ is the minimum of the model, and the inflaton sector
does not contribute to supersymmetry breaking ($F_{\phi} = 0$) at late
times.

Let us first consider the superpotential $W = m_\phi \, \Phi^2 /2\,$.   
Evaluation of eqs.~(\ref{relat}) for the masses of the inflaton
components simply gives
\begin{equation}
m_{\tilde \phi} = {\rm e}^{2-\sqrt{3}} \, m_\phi \;\;\;,\;\;\;
m_{\phi_1\,,\,\phi_2} = {\rm e}^{2-\sqrt{3}} \, \left[ m_\phi \pm
\frac{\mu^2}{M_P}\, \left( 1 - \frac{\sqrt{3}}{2} \right) \right]
\end{equation}
The nonthermal production of inflatinos and gravitinos in this model has been
studied in~\cite{gra2}. The results have been briefly summarized in
section~\ref{sec2} and at the end of the previous subsection.

Let us now consider the superpotential $W = \lambda \, \Phi^3 / 3\,$.  In
this model as well, the  nonperturbative production of inflatinos and
gravitinos is analogous to the ones already discussed. Inflatinos are
mainly produced during the first few oscillations of the inflaton field,
with a typical momentum~\cite{gra1} $k \sim m_\phi \sim \sqrt{\lambda} \,
\phi_0\,$, $\phi_0\sim M_P$ being the initial amplitude of the inflaton
oscillations.  There is however an important difference, due to the fact
that, during the oscillations, in the $V \left( \phi \right) \sim \phi^4$
model the inflaton field has the equation of state of radiation, and its
energy density scales as $R^{-\,4}\,$. Thus the energy density of $\phi$
oscillations is given by
\begin{equation}
\rho_\phi \sim  \lambda \phi_0^4~(R_\phi/R)^4
\label{amp4}
\end{equation}
In this phase both $n_{\tilde \phi}$ and $s \equiv \rho_\phi^{3/4}$ scale
as $R^{-\,3}\,$. The dilution factor then drops out from their ratio (cf.
the last factor in eq.~(\ref{inflab})) and one is simply left with
\begin{equation}
\frac{n_{\tilde \phi}}{s} \sim 10^{-\,2} \, \frac{\left( \sqrt{\lambda} \,
\phi_0 \right)^3}{\left( \lambda \, \phi_0^4 \right)^{3/4}} \sim 10^{-12} \,
\left( \frac{\lambda}{10^{-13}} \right)^{3/4}
\label{chaoinfl2}
\end{equation}
As for the previous cases, nonthermal gravitino production is expected to
occur mainly during the oscillations of the Polonyi field, with a negligible
final abundance.

In the minimum of $V(\phi)$ the masses of the two
inflaton components, as well as that of the inflatino, vanish, although
the inflaton is still massive during its oscillations. Since the amplitude
$\phi$ changes very slowly with respect to the oscillations timescale,
we can get the inflaton mass from the average of $\sqrt{d^2 V/d^2 \phi}$
during one oscillation. Up to a numerical factor of order one, we can set
\begin{equation}
m_\phi \left( t \right) \sim \sqrt{\lambda} \, \phi \left(t \right)
\label{mass4}
\end{equation}
 It is
interesting to note that in this model the inflaton cannot decay
gravitationally, since with the above mass one has 
\begin{equation}
\Gamma_{\rm grav} \simeq \frac{m_\phi^3}{M_P^2} \simeq
\frac{\lambda^{3/2} \phi_0^3}{M_P^{2}}\,
\left( \frac{R_\phi}{R} \right)^3
\end{equation}
This should be compared to the Hubble rate
\begin{equation}
H \sim \frac{\lambda^{1/2} \phi_0^2}{M_P} \, \left(\frac{R_\phi}{R}
\label{h3}
\right)^2
\end{equation}
Clearly the gravitational decay rate is always smaller than
$H$. For a non-gravitational decay with a rate $\Gamma = g^2 \,
m_\phi\,\sim g^2 \lambda^{1/2} \phi_0 (R_\phi/R)$, which should be
compared to eq. (\ref{h3}). Thus the scale factor at decay is
$(R_\phi/R_{d\phi}) \sim M_P g^2/\phi_0$ and the inflaton
lifetime is given by
\begin{equation}
t \sim \frac{M_P^{-\,1}}{\lambda^{1/2}\,g^4} \sim m_{3/2}^{-\,1} \,
\left(
\frac{4 \cdot 10^{-\,3}}{g} \right)^4
\end{equation}
compared here with the timescale of the Polonyi oscillations.
Thus although the inflatino abundance in eq.~(\ref{chaoinfl2}) is
somewhat high, its decay (which is necessarily non-gravitational) does not
lead to gravitino production.

\section{Discussion}

The simple models discussed in this letter were chosen to highlight the
potential problems induced by the inflatino. While more general remarks
have been made in earlier sections, explicit computations have been
performed for a single inflaton supermultiplet along
with the supersymmetry breaking sector described by a Polonyi superfield.
As remarked earlier, we have ignored the cosmological problems associated
directly with the Polonyi scalar's evolution. We have also neglected the
potentially important evolution of other scalar fields or other sources of
entropy production.  

Since the MSSM contains many flat directions, the cosmological evolution 
along these directions can have a direct impact on the issues discussed
here.  For example, it is well known that some flat directions may be
responsible for generating the baryon asymmetry of the Universe
\cite{ad}. In Affleck-Dine baryogenesis scenarios, if the initial field
values of the flat direction is larger than a critical value, $\eta_0
\ga m_{3/2}^{5/12} M_P^{4/3}/m_\phi^{3/4}$, where $\eta$ is the A-D
condensate, the dominant source of entropy comes from $\eta$ decays rather
than inflaton decays, although in this case, the baryon asymmetry is
generally too large. For smaller initial values of $\eta_0$, the A-D
condensate decays during inflaton oscillations. Depending on the value,
$\eta_0$, late entropy production by moduli decay may be a welcome feature
to bring the baryon asymmetry down \cite{cgmo}. 

We have shown that the decay of inflatons to/from inflatinos $+$ gravitino
is potentially a serious problem for a generic inflationary model based on
supergravity. This problem is particularly enhanced when the inflaton
sector is coupled only gravitationally to matter. In this case
the decay channels with a final state gravitino can have rates comparable
to the decays into matter. This existence of these gravitino producing
decays depends on the mass spectrum of the model. In
the particular case of Polonyi-like supersymmetry breaking, we have shown
that such decays are in fact kinematically forbidden, since the mass
difference between the inflaton and the inflatino is smaller than the
gravitino mass.

This fortunate property is not expected to hold for a general
supersymmetry breaking potential. More generally, the direct gravitino
production must be forbidden by imposing a strong hierarchy between the
scale of inflation $\Delta$ and the size of supersymmetry breaking. In
gravitational mediated supersymmetry breaking schemes ($m_{3/2} \sim
100\,$ GeV) this can be translated into an upper bound on $\Delta\,$. This
limit weakens as the initial size of the inflaton oscillations is
decreased. For initial oscillations of the order the Planck scale
(primordial inflation) the bound is quite severe, and a safe value for
the inflationary scale is
$\Delta \ga 10^{13}\,$ GeV. Models of single scale inflation with a scale
dictated by the size of the density fluctuations are typically compatible
with this bound.

\vskip 0.5in
\vbox{
\noindent{ {\bf Acknowledgments} } \\
\noindent
The work of K.A.O. was partially supported by DOE grant
DE--FG02--94ER--40823. The work of H.P.N. an M.P. was supported by the
European Commission RTN programmes HPRN-CT-2000-00148 and 00152. M.P.
would like to thank the Theoretical Physics Institute of the
University of Minnesota for their friendly hospitality during some stages
of this work.}

\end{document}